\documentclass[a4paper,11t]{article}
\usepackage[top=2.54cm, bottom=2.54cm, left=2.54cm, right=2.54cm]{geometry}
\usepackage{amsmath}
\usepackage{amsfonts}
\usepackage{amssymb}
\usepackage{graphicx}
\usepackage{bm}

\usepackage{enumerate}
\usepackage{color}
\usepackage[sort,comma]{natbib}
\usepackage{float}
\usepackage[colorlinks=true, allcolors=blue]{hyperref}
\usepackage{dirtytalk}
\usepackage{pdflscape}
\usepackage{booktabs}
\usepackage{changepage}
\usepackage{relsize}
\usepackage{ulem}
\usepackage{etoolbox}
\usepackage{tikz}
\usepackage{mathtools}
\usepackage{algorithm}
\usepackage{algpseudocode}

\makeatletter
\def\vec{\mathop{\operator@font vec}\nolimits}
\makeatother

\title{Scalable approximation of the transformation-free linear simplicial-simplicial regression via constrained iterative reweighted least squares}
\author{Michail Tsagris$^1$ and Omar Alzeley$^2$ \\
\\
$^1$ Department of Economics, University of Crete, 
Gallos Campus, Rethymnon, Greece \\
\href{mailto:mtsagris@uoc.gr}{mtsagris@uoc.gr} \\
$^2$ Department of Mathematics, Umm Al-Qura University, Saudi Arabia \\
\href{mailto:oazeley@uqu.edu.sa}{oazeley@uqu.edu.sa}
}

\begin{document}

\maketitle

\begin{center}
{\bf Abstract}
\end{center}
Simplicial–simplicial regression concerns statistical modeling scenarios in which both the predictors and the responses are  vectors constrained to lie on the simplex. \cite{fiksel2022} introduced a transformation-free linear regression framework for this setting, wherein the regression coefficients are estimated by minimizing the Kullback–Leibler divergence between the observed and fitted compositions, using an expectation–maximization (EM) algorithm for optimization. In this work, we reformulate the problem as a constrained logistic regression model, in line with the methodological perspective of \cite{tsagris2025}, and we obtain parameter estimates via constrained iteratively reweighted least squares. Simulation results indicate that the proposed procedure substantially improves computational efficiency—yielding speed gains ranging from $6\times--326\times$—while providing estimates that closely approximate those obtained from the EM-based approach. 

\section{Introduction}
Simplicial, or compositional data, data\footnote{In econometrics, these are commonly referred to as multivariate fractional data \citep{mullahy2015,murteira2016}.} are non-negative multivariate vectors whose components convey only relative information. When the vectors are scaled to sum to 1, their sample space is the standard simplex,
\begin{eqnarray} \label{simplex}
\mathbb{S}^{D-1}=\left\lbrace(y_1,\ldots,y_D)^\top \bigg\vert y_i \geq 0,\sum_{i=1}^Dy_i=1\right\rbrace,
\end{eqnarray}
where $D$ denotes the number of components.

Such data arise in a wide range of scientific disciplines, and the extensive literature on their proper statistical treatment attests to their prevalence in practical applications.\footnote{For numerous examples of real-world applications involving simplicial data see \cite{tsagris2020}.}

The frequent appearance of simplicial variables in regression settings has motivated the development of appropriate regression methodologies, leading to several recent methodological advances. Most contributions in this field focus either on the case of a simplicial response with real-valued predictors (simplicial--real regression) or on the converse setting with simplicial predictors and real-valued responses (real--simplicial regression). By contrast, the simplicial--simplicial regression setting, in which both the response and predictor variables are simplicial, has received comparatively limited attention. This constitutes the main focus of the present work.

Applications involving simplicial responses and simplicial predictors include, among others, \cite{wang2013}, who modeled the relationship between economic outputs and inputs in China, and \cite{dimarzio2015}, who predicted the composition of the moss layer using the composition of the O-horizon layer. \cite{chen2017} investigated the association between age structure and consumption structure across economic regions. \cite{filzmoser2018} studied differences in educational composition across countries. \cite{ait2003} and \cite{alenazi2019} examined relationships between alternative methods of estimating white blood cell type compositions. \cite{chen2021} explored associations between chemical metabolites of \textit{Astragali Radix} and plasma metabolites in rats following administration. \cite{tsagris2025} analyzed relationships between crop production and cultivated area in Greece, as well as voting proportion dynamics in Spain. Finally, \cite{rios2025} investigated associations between high-dimensional multi-omics simplicial datasets.

Most existing approaches to simplicial--simplicial regression apply transformations to both simplicial variables. For example, \cite{hron2012}, \cite{wang2013}, \cite{chen2017}, and \cite{han2022} employed log-ratio transformations for both predictors and responses followed by multivariate linear regression. \cite{alenazi2019} transformed simplicial predictors via the $\alpha$-transformation \citep{tsagris2011}, applied principal component analysis, and subsequently fitted a Kullback--Leibler divergence (KLD) regression (multinomial logit) model \citep{murteira2016}. In contrast, \cite{fiksel2022} proposed a transformation-free linear regression (TFLR) framework, in which the regression coefficients lie on the simplex and are estimated by minimizing the KLD between observed and fitted simplicial responses using an Expectation--Maximization (EM) algorithm.

A limitation of the EM-based estimation procedure for the TFLR model is its computational burden, as the EM algorithm is generally slow to converge. To alleviate this issue, we employ a constrained iteratively reweighted least squares (CIRLS) algorithm, incorporating simplex constraints on the regression coefficients. The implementation of the EM algorithm in \cite{tsagris2025} was shown to be approximately four times faster than that of \cite{fiksel2022}. Our simulation studies demonstrate that CIRLS achieves additional substantial computational gains, yielding speed improvements ranging from $2\times$ to $255\times$, depending on the scenario, while producing estimates that closely approximate those obtained via the EM algorithm. These computational benefits are particularly relevant in contexts involving: (a) analysis of multiple datasets, (b) large-scale simulation studies, (c) high-dimensional compositions \citep{rios2025}, and (d) permutation- or bootstrap-based inference procedures \citep{fiksel2022,tsagris2025}.

The next section presents the TFLR model and the EM algorithm used for parameter estimation. Section~\ref{sec:cirls} describes the CIRLS algorithm for logistic regression and its adaptation to the TFLR framework. Section~\ref{sec:complexity} discusses the computational complexity of both approaches, Section~\ref{sec:sims} reports simulation results, and the final section provides concluding remarks.

\section{The TFLR model} \label{sec:tflr}
\cite{fiksel2022} developed the TFLR model as a novel framework for modeling simplicial data, addressing methodological limitations inherent in existing approaches.

The TFLR model is defined as
\begin{align}
\mathbb{E}[\bm{Y}|\bm{X}] = \bm{X}\bm{B},
\end{align}
where $\bm{X} \in \mathcal{S}^{D_p-1}$ denotes a $D_p$-dimensional simplicial predictor and $\bm{Y} \in \mathcal{S}^{D_r-1}$ denotes a $D_r$-dimensional simplicial response. The model relates the $k$-th component of the simplicial response to the predictor through the linear mapping
\begin{eqnarray} \label{lt}
E(\bm{Y}_k \mid \bm{X}) = \sum_{j=1}^{D_p} \bm{X}_j B_{jk},
\end{eqnarray}
where the coefficient matrix $\bm{B}$ itself lies in a product of simplices,
\[
\bm{B} \in \left\lbrace \mathbb{R}^{D_p \times D_r} \mid B_{jk} \geq 0,\; \sum_{k=1}^{D_r} B_{jk} = 1 \right\rbrace.
\]
The elements of $\bm{B}$ are estimated by minimizing the Kullback--Leibler divergence between the observed and fitted response compositions:
\begin{eqnarray} \label{kld}
\min_{\bm{B}} \left\lbrace \sum_{i=1}^n\sum_{k=1}^{D_r}
y_{ik}\log\left(\frac{y_{ik}}{\sum_{j=1}^{D_p}x_{ij}B_{jk}}\right)\right\rbrace
\;=\;
\max_{\bm{B}}
\left\lbrace \sum_{i=1}^n\sum_{k=1}^{D_r}
y_{ik}\log\left(\sum_{j=1}^{D_p}x_{ij}B_{jk}\right) \right\rbrace .
\end{eqnarray}

Unlike conventional approaches relying on log-ratio transformations (ALR, CLR, ILR), the TFLR model operates directly in the simplex, thereby avoiding interpretability challenges and distributional distortions introduced by transformations. The associated estimation procedure is based on first-moment specifications via estimating equations rather than full likelihood specification, providing robustness to a variety of data-generating mechanisms. Furthermore, the method naturally accommodates zeros and ones in both predictors and responses, overcoming a major limitation of log-ratio-based techniques, which typically require preprocessing such as zero-replacement.

However, as noted by \cite{tsagris2025}, this methodology generally allows for zeros in the response but does not fully accommodate zeros in the simplicial predictors. Consider, for example, an observation $\bm{x}_i = (x_{i1}, x_{i2}, 0, 0)$, where $x_{i1}$ and $x_{i2}$ are positive, and suppose that the $k$-th column of the estimated matrix $\bm{B}$ is $(0, 0, B_{3k}, B_{4k})^\top$ with $B_{3k}$ and $B_{4k}$ strictly positive. Alternatively, a column of $\bm{B}$ may consist entirely of zeros. In both situations, the product $x_{ik}B_{ik} = 0$, rendering (\ref{kld}) undefined. To circumvent this issue, \cite{fiksel2022} introduce a small positive constant ($\delta = 10^{-8}$) when computing the KLD.

\subsection{EM algorithm}
\cite{fiksel2022} employed the EM algorithm to estimate the parameters of the TFLR model, and in particular the column-stochastic transition matrix $\bm{B}$. The estimation problem is formulated as a missing-data problem, wherein each observed simplicial vector $\bm{y}_i$ is interpreted as the outcome of a two-stage generative mechanism. Specifically: 
(a) for each component $j$ of $\bm{x}_i$, a component $k$ of $\bm{y}_i$ is selected with probability $b_{jk}$; and 
(b) the contributions from all components of $\bm{x}_i$ are subsequently aggregated according to their proportions $x_{ij}$. 
The steps of the EM algorithm are outlined below.

\begin{enumerate}
\item \textbf{Initialization:} Set $\bm{B}^{(0)}$ to a valid column-stochastic matrix, i.e., each row is equal to $1/D_p$.
\item \textbf{Expectation step (E-step):} For iteration $t$, compute the expected latent allocations:
\begin{eqnarray*}
z_{ijk}^{(t)} = \frac{x_{ij}B_{jk}^{(t-1)}}{\sum_{j'=1}^p x_{ij}B_{jk}^{(t-1)}} \cdot y_{ik}
\end{eqnarray*}
where $z_{ijk}$ represents the contribution of predictor component $j$ to response component $k$ for observation $i$.

\item \textbf{Maximization step (M-step):} Update the transition matrix elements:
\begin{eqnarray*}  
B_{jk}^{(t)} = \frac{\sum_{i=1}^n z_{ijk}^{(t)}}{\sum_{k=1}^q \sum_{i=1}^n z_{ijk}^{(t)}}
\end{eqnarray*}

\item \textbf{Convergence check:} Repeat steps 2-3 until a convergence criterion is met:
\begin{eqnarray*}
\|\bm{B}^{(t)} - \bm{B}^{(t-1)} \|_1 < \epsilon  \ \ \text{or} \ \ \text{KLd}^{(t-1)} - \text{KLD}^{(t)} < \epsilon,
\end{eqnarray*}
where $\|.\|_1$ denotes the $L_1$ norm and $\epsilon$ is a small tolerance value.
\end{enumerate}

The EM updates naturally satisfy the non-negativity and sum-to-one constraints on each row of $\bm{B}$. Moreover, each iteration of the algorithm is guaranteed to increase the (pseudo-)likelihood function. Numerical underflow and overflow issues are mitigated by operating on log-transformed probabilities when necessary, and the algorithm accommodates zeros in the data (particularly in the simplicial response) without requiring zero-value imputation.

In addition, the Initialization Step employs starting values derived from the simplicially constrained least squares estimator of \cite{tsagris2025}. This enhancement yielded an implementation that was reported to be approximately four times faster than that of the \textsf{R} package \textsf{codalm} \citep{codalm2021}, albeit at the cost of increased memory requirements.

\section{TFLR with CIRLS} \label{sec:cirls}
While the EM algorithm provides a convenient latent-allocation interpretation of the TFLR model, it is not essential for estimating the coefficient matrix $\bm{B}$. The EM procedure maximizes the KLD objective in (\ref{kld}), up to an additive constant, indirectly by introducing latent variables and iteratively improving a lower bound of the KLD criterion.

An alternative perspective arises by noting that the fitted compositions
\[
\hat{y}_{ik}(\bm{B}) = \sum_{j} x_{ij} B_{jk}
\]
behave as probabilities. Consequently, the optimization problem can be viewed as a constrained regression problem defined on the simplex. This viewpoint motivates the use of iteratively reweighted least squares (IRLS), which replaces the EM lower bound with a quadratic surrogate obtained via a second-order Taylor expansion of the objective function.

Thus, both EM and IRLS yield weighted updates for $\bm{B}$, although IRLS generates Newton-type additive steps that typically require substantially fewer iterations. Each IRLS update reduces to solving a strictly convex quadratic program subject to simplicial constraints, leading to a computationally efficient approximation to the EM solution while preserving the structure of the parameter space.

In what follows, we present preliminary concepts regarding the TFLR model, the IRLS procedure, and its constrained extension (CIRLS), and demonstrate how the latter can be applied to estimate the regression coefficients of the TFLR framework.

\subsection{Iteratively reweighted least squares for logistic regression}
The logistic regression is a generalised linear model where the response variable follows a Bernoulli distribution $y_i \sim \text{Bernoulli}(p_i)$, the ordinary link function is the logit $g(p_i) = \text{logit}(p_i) = \ln\left(\frac{p_i}{1-p_i}\right)$, the mean function is $\mathbb{E}[y_i] = p_i$, the variance function is $\text{Var}[y_i] = p_i(1-p_i)$, and the model is expressed as 
\begin{equation*}
\ln\left(\frac{p_i}{1-p_i}\right) = \bm{x}_i^T\bm{\beta} \Leftrightarrow
p_i = \frac{\exp(\bm{x}_i^T\bm{\beta})}{1 + \exp(\bm{x}_i^T\bm{\beta})} 
= \frac{1}{1 + \exp(-\bm{x}_i^T\bm{\beta})},
\end{equation*}
where $\bm{x}_i$ is a vector containing the values of the predictor variables for the $i$-th observation.

For $n$ independent observations, the log-likelihood function is
\begin{equation} \label{ell}
\ell(\bm{\beta}) = \sum_{i=1}^n \left[ y_i \ln(p_i) + (1-y_i)\ln(1-p_i) \right] \\
= \sum_{i=1}^n \left[ y_i \bm{x}_i^T\bm{\beta} - \ln(1 + \exp(\bm{x}_i^T\bm{\beta})) \right]
\end{equation}

To estimate the regression coefficients $\bm{\beta}$), the Newton-Raphson algorithm that uses the first and second order derivatives may be employed. The score function (gradient of log-likelihood) is
\begin{equation*}
\bm{s}(\bm{\beta}) = \frac{\partial \ell}{\partial \bm{\beta}} = \sum_{i=1}^n \bm{x}_i(y_i - p_i) = \bm{X}^T(\bm{y} - \bm{p})
\end{equation*}

The Fisher information matrix (negative expected Hessian of log-likelihood) is
\begin{equation*}
\bm{I}(\bm{\beta}) = -\mathbb{E}\left[\frac{\partial^2 \ell}{\partial \bm{\beta} \partial \bm{\beta}^T}\right] = \sum_{i=1}^n p_i(1-p_i)\bm{x}_i\bm{x}_i^T = \bm{X}^T\bm{W}\bm{X},
\end{equation*}
where $\bm{W} = \text{diag}(w_1, \ldots, w_n)$ is the weight matrix, where $w_i= p_i(1-p_i)$.

For logistic regression, the iteratively reweighted least squares (IRLS) algorithm works by updating the $\beta_s$ in an iterative way. Define $z_i$ to be the working response
\begin{equation*}
z_i = \bm{x}_i^T\bm{\beta} + \frac{y_i - p_i}{w_i}.
\end{equation*}
The update scheme of the $\beta_s$ comes from the fact that at each iteration, the $\bm{\beta}^{t+1}$ is produced by an ordinary least squares (OLS) minimization problem
\begin{equation*}
\bm{\beta}^{(t+1)} = \min_{\bm{\beta}}\sum_{i=1}^nw_i\left(z_i-\bm{x}_i^\top\bm{\beta}\right)^2 = \min_{\bm{\beta}}\left[\left(\bm{z}-\bm{X}^\top\bm{\beta}\right)^\top\bm{W}\left(\bm{z}-\bm{X}^\top\bm{\beta}\right)\right].
\end{equation*}
Hence, the updated $\bm{\beta}$ is given by\footnote{This expression explicitly shows that IRLS for logistic regression is equivalent to Newton-Raphson optimization of the log-likelihood function.}.
\begin{equation} \label{be}
\bm{\beta}^{(t+1)} = (\bm{X}^T\bm{W}^{(t)}\bm{X})^{-1}\bm{X}^T\bm{z}^{(t)}.
\end{equation}

\subsection{Logistic regression with identity link and simplicial constraints}
Suppose we wish to maximize Eq. (\ref{ell}), using the identity link, $\mathbb{E}[y_i] = \bm{x}_i^\top\bm{\beta}$, subject to the following simplex constraints: $0 \leq \beta_j \leq 1$, for all $j = 1, 2, \ldots, p$ and $\sum_{j=1}^p \beta_j = 1$. Note, that in this formulation, we have assumed there is no constant. Two notable differences occur, the working response $\bm{z}$ equals the original response vector $\bm{y}$, $z_i=y_i$ and $w_i=\left[(p_i(1-p_i)\right]^{-1}$.

This approach has been used, for example, to estimate predator diet within the quantitative fatty acid signature analysis framework \citep{iverson2004}. However, the standard IRLS algorithm must be modified to accommodate simplex constraints. In particular, at each iteration, after computing the unconstrained update, one would need to project the estimate back onto the feasible region defined by the simplex. Such projection-based corrections, however, may not be optimal from a likelihood-based perspective. A more principled alternative is to formulate the constrained weighted least squares problem at each IRLS iteration as a quadratic programming (QP) problem, leading to the constrained IRLS (CIRLS) formulation \citep{masselot2025}.

At iteration $t$ of the IRLS algorithm, we need to solve the following constrained weighted least squares problem
\begin{eqnarray*}
\min_{\bm{\beta}} \quad & \sum_{i=1}^n w_i^{(t)} (z_i^{(t)} - \bm{x}_i^T \bm{\beta})^2 \\
\text{subject to} \quad & \sum_{j=1}^p \beta_j = 1 \\
& 0 \leq \beta_j \leq 1, \quad j = 1, 2, \ldots, p.
\end{eqnarray*}

This can be rewritten in standard QP form as:
\begin{eqnarray*}
\min_{\bm{\beta}} \quad & - \bm{d}^T \bm{\beta} + \frac{1}{2}\bm{\beta}^T \bm{D} \bm{\beta}\\
\text{subject to} \quad & \bm{A} \bm{\beta} \geq \bm{b},
\end{eqnarray*}
where:
\begin{eqnarray*}
\bm{D} &=& \bm{X}^T\bm{W}^{(t)}\bm{X} \\
\bm{d} &=& \bm{X}^T\bm{W}^{(t)}\bm{z}^{(t)} \\
\bm{A} &=& \left(\begin{array}{c}
\bm{j}_p^\top \\
\bm{I}_p 
\end{array}\right)  \\
\bm{b} &=& (1, \bm{j}_p^\top)^\top,
\end{eqnarray*}
where $\bm{j}_p=(1,\ldots,1)^\top$ is the p-dimensional vector of $1s$, and $\bm{I}_p$ is the p-dimensional identity matrix. The first row of $\bm{A}$ corresponds to the sum of the $\beta_s$ that must be equal to 1, which is denoted by the first element of the $\bm{b}$ vector. Hence the $p + 1$ constraints, are in fact one equality constraint and $p$ inequality constraints. 

\begin{algorithm}
\caption{IRLS for logistic regression with identity link function and simplex constraints}
\begin{algorithmic}[1]
\State \textbf{Initialize:} $\bm{\beta}^{(0)}$ satisfying the simplex constraints
\State $t \gets 0$
\State Define $\bm{A}$ and $\bm{b}$
\While{not converged}
    \State $\bm{\mu}^{(t)} \gets \bm{X}\bm{\beta}^{(t)}$  \Comment{Mean equals linear predictor for identity link}
    \State Compute weights: $w_i^{(t)} \gets \left[\mu_i^{(t)}(1-\mu_i^{(t)})\right]^{-1}$
    \State $\bm{W}^{(t)} \gets \text{diag}(w_1^{(t)}, \ldots, w_n^{(t)})$
    \State Compute working response: $\bm{z}^{(t)} \gets \bm{y}$  \Comment{For identity link}
    \State Set up QP problem:
    \State \quad $\bm{D} \gets \bm{X}^T\bm{W}^{(t)}\bm{X}$
    \State \quad $\bm{d} \gets \bm{X}^T\bm{W}^{(t)}\bm{z}^{(t)}$
    \State Solve QP: $\bm{\beta}^{(t+1)} \gets \arg\min_{\bm{\beta}} \frac{1}{2}\bm{\beta}^T \bm{D} \bm{\beta} - \bm{d}^T \bm{\beta}$ subject to the simplicial constraints
    \State $t \gets t + 1$
\EndWhile
\State \Return $\bm{\beta}^{(t)}$
\end{algorithmic}
\end{algorithm}

The QP approach offers several advantages over alternative methods for handling simplex constraints. First, the constraints are explicitly incorporated into the optimization problem at each iteration, rather than being imposed post hoc following an unconstrained update. Moreover, QP solvers employ specialized algorithms designed to efficiently manage constraints while maintaining numerical stability. For convex optimization problems---such as the weighted least squares problem arising at each IRLS iteration---QP solvers guarantee convergence to the global optimum under the imposed constraints. In contrast to transformation-based approaches (e.g., softmax reparametrization), QP directly operates on the original parameters, thereby avoiding potential numerical difficulties associated with reparametrization. Finally, modern QP solvers are highly optimized and capable of handling large-scale problems with considerable computational efficiency.

A valid starting point for $\bm{\beta}^{(0)}$ that satisfies the simplex constraints could be
\begin{eqnarray*}
\beta_j^{(0)} = \frac{1}{p} \quad \text{for } j = 1, 2, \ldots, p.
\end{eqnarray*}

This approach assigns uniform weights to each regression coefficient. As an alternative initialization strategy, starting values can be obtained by fitting ordinary least squares under simplicial constraints on the coefficient vector. We employ an absolute difference convergence criterion based on successive log-likelihood values. Under certain parametric configurations, the optimal solution may reside on the boundary of the simplex, resulting in one or more coefficients attaining their boundary values of 0 or 1. Such boundary solutions can induce degeneracy in the optimization problem. While quadratic programming solvers are typically equipped to handle degenerate cases, implementation with solvers employing interior point methods or alternative algorithms designed for robust handling of degeneracy may prove more reliable and computationally stable.

\subsection{CIRLS formulation of the TFLR}
To formulate the TFLR as a univariate logistic regression model we follow \cite{tsagris2025}. They constructed the $\bm{D}$ matrix is a $D_{rp} \times D_{rp}$ diagonal matrix, where $D_{rp}=D_r \times D_p$, and is related to the $\bm{X}^\top \bm{X}$ in the following manner
\begin{eqnarray} \label{Dmat}
\bm{D} =
\left( \begin{array}{cccc}
\bm{X}^\top\bm{V}_1\bm{X} &  \bm{0} & \ldots & \bm{0}\\
\bm{0} & \ddots & \bm{0} & \vdots \\
\vdots & \bm{0} & \ddots &  \bm{0} \\
\bm{0} & \ldots & \bm{0} &  \bm{X}^\top  \bm{V}_{D_r}\bm{X} 
\end{array} \right),
\end{eqnarray}
where $V_k=\text{diag}\left(W_{1k}, \ldots W_{nk}\right)$ is a diagonal matrix containing the $k$-column of the weighting matrix $\bm{W}$, and $\bm{I}_{D_r}$ is the $D_r \times D_r$ identity matrix. The vector $\bm{d}$ is 
given by the vectorization of the product between $\bm{X}$ and $\bm{W} \cdot \bm{Y}$, 
\begin{eqnarray} \label{dvec}
\bm{d}=\vec\left(\bm{X}^\top\bm{W} \cdot \bm{Y}\right),
\end{eqnarray}
where "$\cdot$" implies the elementwise multiplication of the two matrices. 

Notably, we do not vectorize the simplicial responses or the simplicial predictors, nor the weighting matrix $\bm{W}$. The matrix $\bm{B}$ is obtained via quadratic programming, from which we subsequently compute the fitted response matrix and the corresponding KLD value. The weighting matrix is then updated, and this iterative process continues until the KLD converges. Initial values for $\bm{B}$ are obtained using the constrained least squares method of \cite{tsagris2025}. 

The construction of matrix $\bm{D}$ (\ref{Dmat}) and vector $\bm{d}$ (\ref{dvec}) imposes substantial memory requirements. Unfortunately, no straightforward approach exists to simultaneously reduce memory demands and parallelize the quadratic programming step—a critical consideration when analyzing high-dimensional simplicial data. 

Solutions obtained via TFLR may reside on the boundary of the simplex, with certain coefficients attaining their limiting values of exactly $0$ or $1$. Under such circumstances, the active-set approach of \cite{goldfarb1983} proves particularly well suited, as it handles degeneracy by maintaining a working set of active constraints whose feasibility is rigorously enforced at each iteration. This mechanism ensures that IRLS updates remain within the simplex and that convergence is not compromised by rank deficiencies in the reduced Hessian.

We employ the dual active-set implementation provided by the \textit{R} package \textsf{quadprog}~\citep{quadprog2019}, which exhibits high computational efficiency when the number of active constraints at the optimum is small—a scenario frequently observed empirically when TFLR solutions lie near the interior of the simplex. While alternative interior-point methods could be deployed, particularly in high-dimensional regimes where the $O(D_p^3)$ factorization cost becomes dominant, we find the active-set method to be both numerically stable and highly effective across the dimensionalities considered in this work.

\section{Computational complexity of EM and CIRLS algorithms} \label{sec:complexity}
We analyze the computational and memory complexity of both the EM and CIRLS algorithms to understand 
their scalability properties and explain the observed speed-up factors.

\subsection{Per-iteration Time Complexity}
\begin{itemize}
\item EM Algorithm: The E-step computes the latent allocations $z_{ijk}^{(t)}$ for all $i \in n$, $j \in D_p$, 
and $k \in D_r$. For each triplet $(i,j,k)$, the E-step costs $\mathcal{O}(n D_p^2 D_r)$. The M-step performs normalization in $\mathcal{O}(n D_p D_r)$ time. Thus, the complexity of the $t$-th iteration is $T_{\text{EM}}^{(t)} = \mathcal{O}(n D_p^2 D_r)$.
\item CIRLS Algorithm: After vectorization, the main computational steps are: a) computing the linear predictor 
$\boldsymbol{\mu}^{(t)} = \tilde{\mathbf{X}} \tilde{\boldsymbol{\beta}}^{(t)}$ in 
$\mathcal{O}(n D_p D_r)$ time; b) computing the weight matrix $\mathbf{W}^{(t)}$ in 
$\mathcal{O}(n D_r)$ time; c) forming $\mathbf{D} = \tilde{\mathbf{X}}^\top \mathbf{W}^{(t)} \tilde{\mathbf{X}}$, 
which is block diagonal with $D_r$ blocks of size $D_p \times D_p$, requiring 
$\mathcal{O}(n D_p^2 D_r)$ operations; and d) solving the QP problem with $D_{rp}$ variables and $\mathcal{O}(D_{rp})$ constraints. Using interior-point methods, the QP solve 
costs $\mathcal{O}(D_{rp}^3) = \mathcal{O}(D_p^3 D_r^3)$. The complexity of the $t$-th iteration is $T_{\text{CIRLS}}^{(t)} = \mathcal{O}(n D_p^2 D_r + D_p^3 D_r^3)$.
\end{itemize}

For typical problems where $n \gg D_p D_r$, both algorithms have similar per-iteration complexity 
$\mathcal{O}(n D_p^2 D_r)$, as the matrix operations dominate the QP overhead. In this regime, the speed-up of CIRLS over EM arises primarily from requiring fewer iterations to converge (typically $K_{\text{CIRLS}} \approx \frac{1}{2}K_{\text{EM}}$ to $\frac{1}{5}K_{\text{EM}}$). For problems with large $D_p \times D_r$ relative to $n$, the QP solving becomes the computational  bottleneck in CIRLS. However, this can be mitigated by exploiting the fact that the $\mathbf{B}$ matrix has $D_r$ independent rows (each satisfying simplex constraints), allowing decomposition into $D_r$ separate QP problems of size $D_p$ that can be solved in parallel.

If EM requires $K_{\text{EM}}$ iterations and CIRLS requires $K_{\text{CIRLS}}$ iterations 
to converge, the total complexities are
$T_{\text{EM}} = \mathcal{O}(K_{\text{EM}} \cdot n D_p^2 D_r)$ and 
$T_{\text{CIRLS}} = \mathcal{O}(K_{\text{CIRLS}} \cdot [n D_p^2 D_r + D_p^3 D_r^3])$.

\subsection{Memory Complexity}
\begin{itemize}
\item EM Algorithm: The main memory requirements are: a) the data matrices $\mathbf{X}$ 
and $\mathbf{Y}$, requiring $\mathcal{O}(n(D_p + D_r))$ space; b) the coefficient matrix 
$\mathbf{B}$, requiring $\mathcal{O}(D_p D_r)$ space; and c) the latent allocation array 
$z_{ijk}$, requiring $\mathcal{O}(n D_p D_r)$ space. Thus,
$M_{\text{EM}} = \mathcal{O}(n D_p D_r)$.

\item CIRLS Algorithm: In addition to the data and coefficients, CIRLS requires: a) the  block-diagonal matrix $\mathbf{D}$, which can be stored efficiently as $D_r$ blocks of size $D_p \times D_p$, requiring $\mathcal{O}(D_p^2 D_r)$ space; b) the QP solver workspace, typically requiring $\mathcal{O}(D_p^2 D_r^2)$ space for storing constraint matrices and factorizations. Thus,
$M_{\text{CIRLS}} = \mathcal{O}(n D_p D_r + D_p^2 D_r^2)$.
\end{itemize}

Table \ref{complexity} summarizes the key results. Based on the complexity analysis and empirical results, with small to medium problems ($n \leq 50{,}000$, $D_p D_r \leq 100$): CIRLS is strongly recommended, offering 5--85$\times$ speed-up with identical accuracy. With large sample sizes, ($n > 50{,}000$): CIRLS remains faster but with diminishing speed-up (2--10$\times$) due to different scaling rates. For typical problems with small to moderate $D_p$ and $D_r$, the memory requirements are similar. EM is more memory-efficient when $D_p \times D_r$ is large.

\begin{table}[ht]
\centering
\caption{Computational and memory complexity comparison of EM and CIRLS algorithms.}
\label{complexity}
\begin{tabular}{lll}
\toprule
& \textbf{EM} & \textbf{CIRLS} \\
\midrule
Per-iteration time & $\mathcal{O}(n D_p^2 D_r)$ & $\mathcal{O}(n D_p^2 D_r + D_p^3 D_r^3)$ \\
Iterations to converge & $K_{\text{EM}}$ & $K_{\text{CIRLS}} < K_{\text{EM}}$ \\
Total time & $\mathcal{O}(K_{\text{EM}} n D_p^2 D_r)$ & $\mathcal{O}(K_{\text{CIRLS}}[n D_p^2 D_r + D_p^3 D_r^3])$ \\
Memory & $\mathcal{O}(n D_p D_r)$ & $\mathcal{O}(n D_p D_r + D_p^2 D_r^2)$ \\
\bottomrule
\end{tabular}
\label{tab:complexity}
\end{table}

\subsection{Implementation details in \textit{R}}
Table \ref{tab:complexity} presents the theoretical complexities of the two algorithms. The bottleneck of the CIRLS is the computation of the block matrix $\bm{D}$ (\ref{Dmat}) that involves performing $2 \times D_r$ cross-product matrix multiplications. Second, computational optimizations have been applied. Instead of computing the full KLD, at each iteration of either algorithm, only the parts involving the fitted values are computed, i.e. the right hand side of Eq. (\ref{kld}) is computed. The matrix $\bm{D}$ (\ref{Dmat}), required by QP, is initialised once, and its elements are updated at every iteration.  

Both algorithms are available in the \textit{R} package \textsf{Compositional} \citep{compositional2025}. The relevant functions are \texttt{tflr()} and \texttt{tflr.irls()} and they return the same output, the running time, the number of iterations required, the KLD value, the estimated regression coefficients and optionally, the fitted values for new observations. 

\section{Simulation studies} \label{sec:sims}
Simulation studies\footnote{The experiments were conducted utilizing a Dell laptop equipped with Intel Core i5-1035G1 CPU at 1GHz, with 256 GB SSD, 8 GB RAM and Windows installed.} were conducted to compare the computational cost (running time) of the EM and the CIRLS estimation procedures for the regression coefficients of the TFLR. For all cases, both simplicial responses and predictors were generated from a Dirichlet distribution. The sample sizes considered were 1,000 up to 10,000 with an increasing step size of 1,000 and from then the step size increased to 5,000 up to 50,000. For all cases, the tolerance value to check convergence was $\epsilon=10^{-6}$, and the results were averaged over $100$ replicates for each combination of dimensionality and sample size. Two scenarios were considered: a) the responses and the predictor variables are independent and b) they are linearly dependent, with the following configurations: 
\begin{itemize}
\item The simplicial response consisted of $D_r = (3, 5, 7, 10)$ components, while the simplicial predictor contained $D_p=(5, 10)$ components. 
\item The simplicial response consisted of $D_r = (20, 30, 40, 50)$ components, while the simplicial predictor contained $D_p=(15, 20)$ components. 
\end{itemize}

EM-1 refers to the implementation in the package \textsf{codalm}, EM-2 refers to the implementation in the package \textsf{Compositional}. The tolerance value to terminate the EM and CIRLS algorithms was set to $10^{-8}$. The simulations were performed in a desktop with Intel Core i9-14900K CPU at 3.2GHz, 128 GB RAM and Windows 11 Pro installed.
 
\subsection{Speed-up factors}
\begin{figure}[!ht]
\centering
\begin{tabular}{cc}
\includegraphics[scale = 0.3]{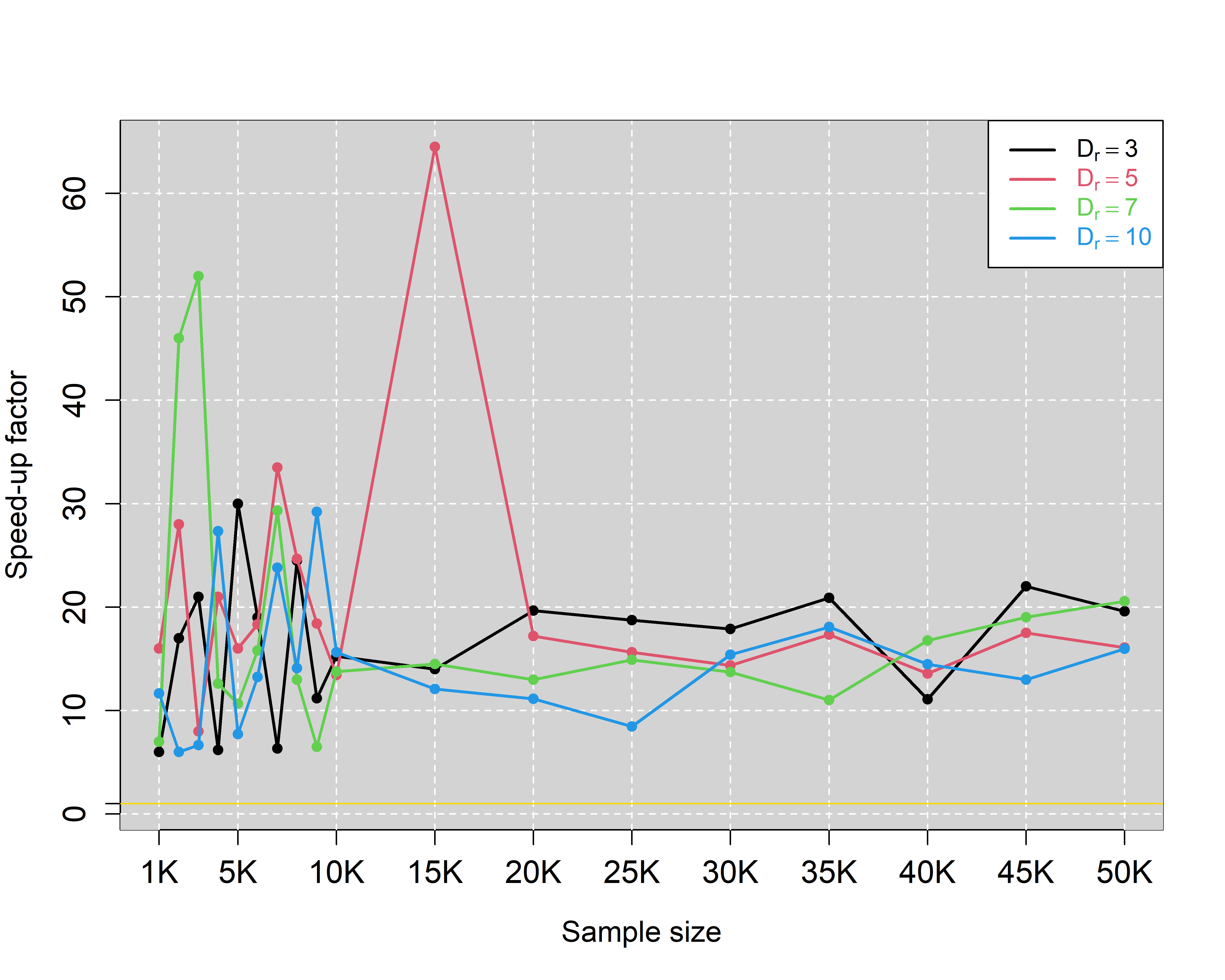}  &
\includegraphics[scale = 0.3]{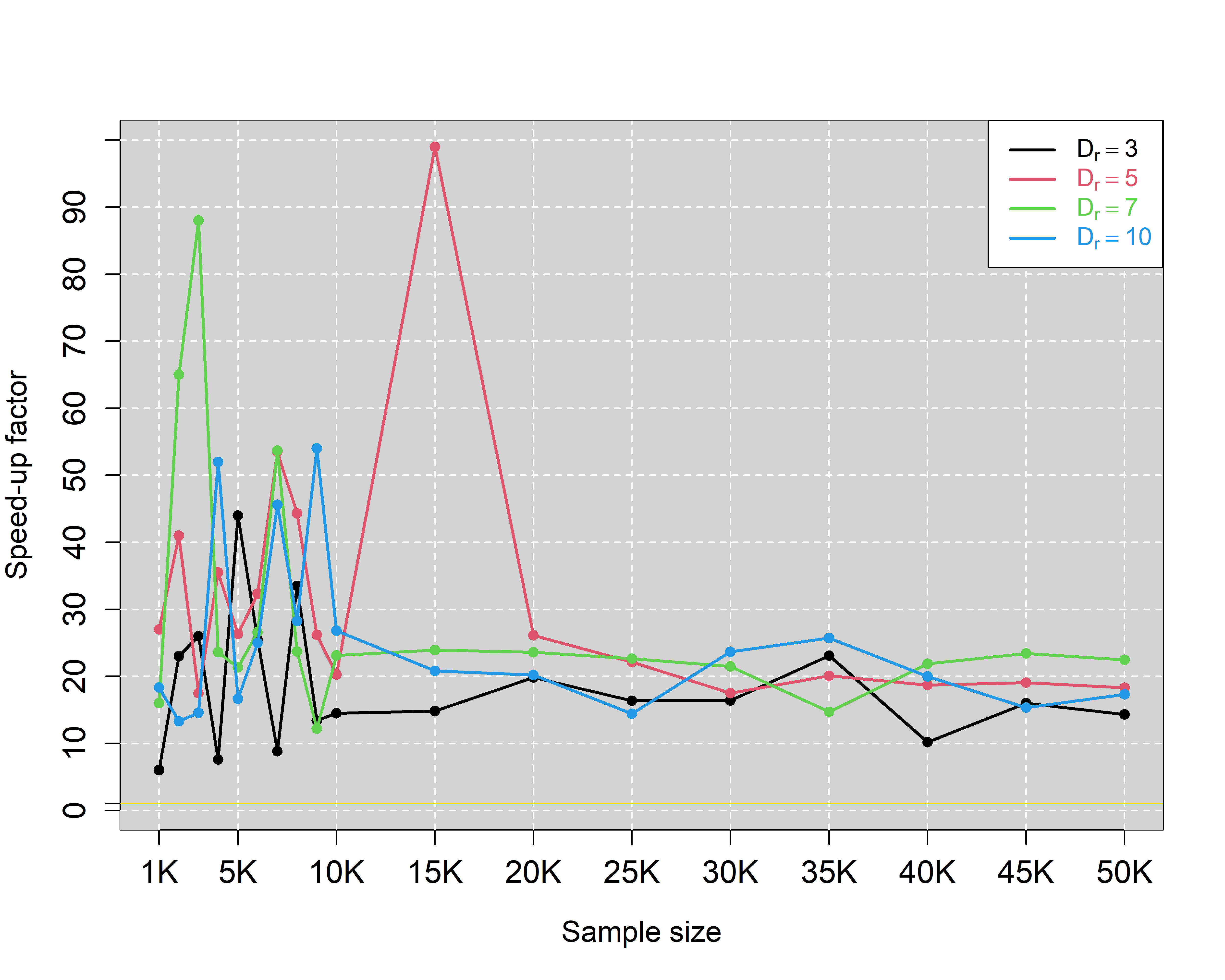}  \\
(a) EM-1 to CIRLS when $p = 5$  &  (b) EM-2 to CIRLS when $p=5$ \\
\includegraphics[scale = 0.3]{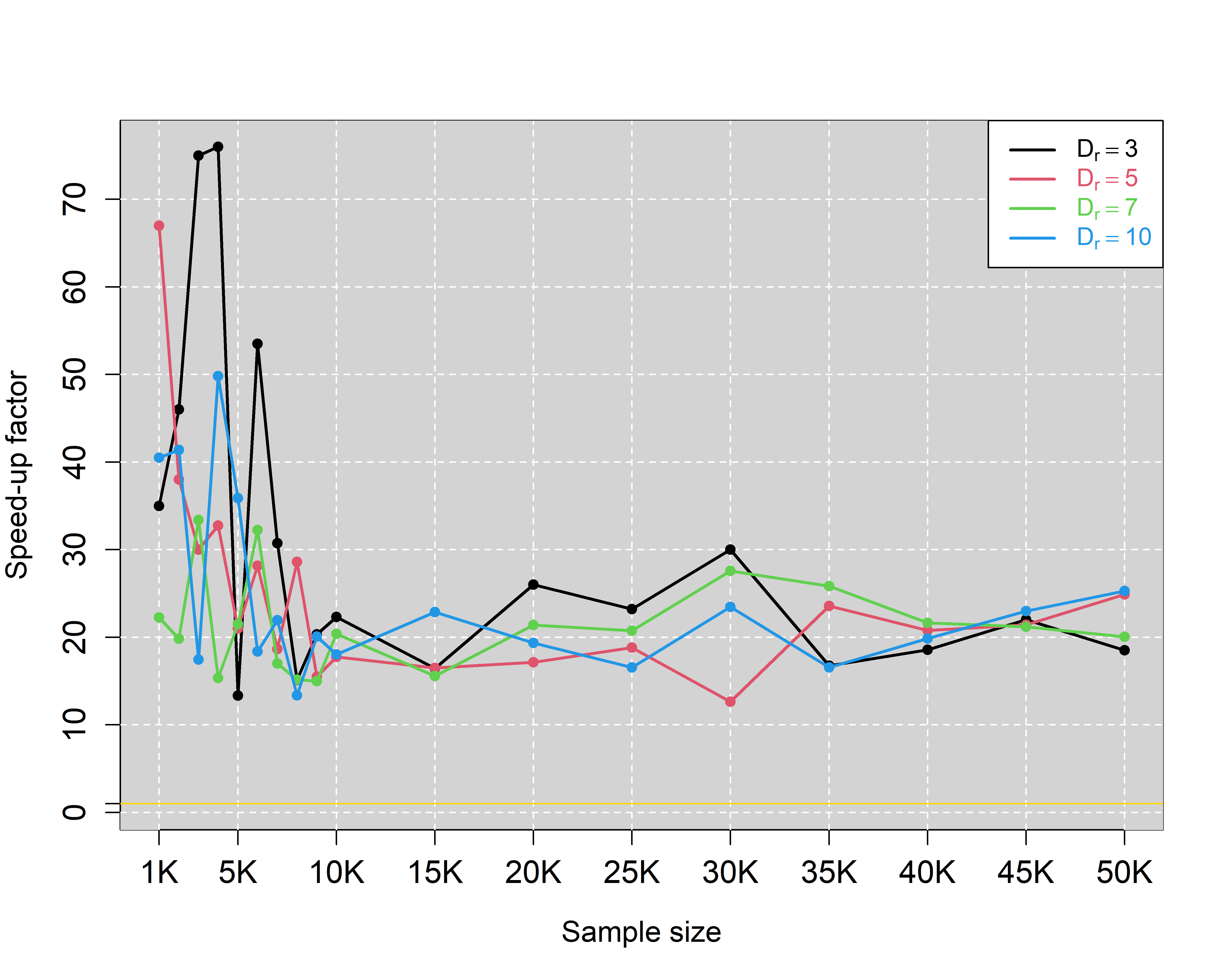}  &
\includegraphics[scale = 0.3]{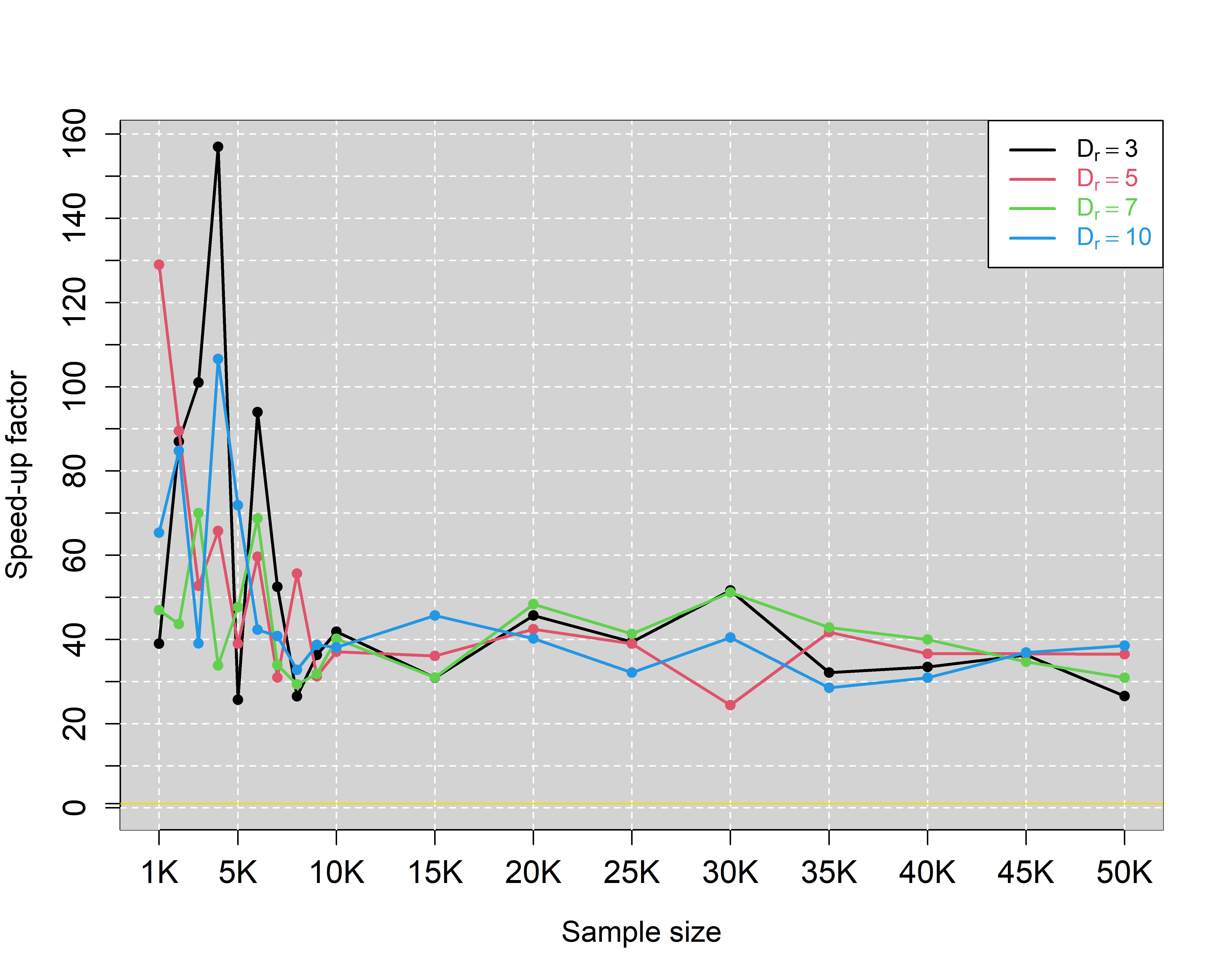}  \\
(c) EM-1 to CIRLS when $p = 10$  &  (d) EM-2 to CIRLS when $p=10$ \\
\includegraphics[scale = 0.3]{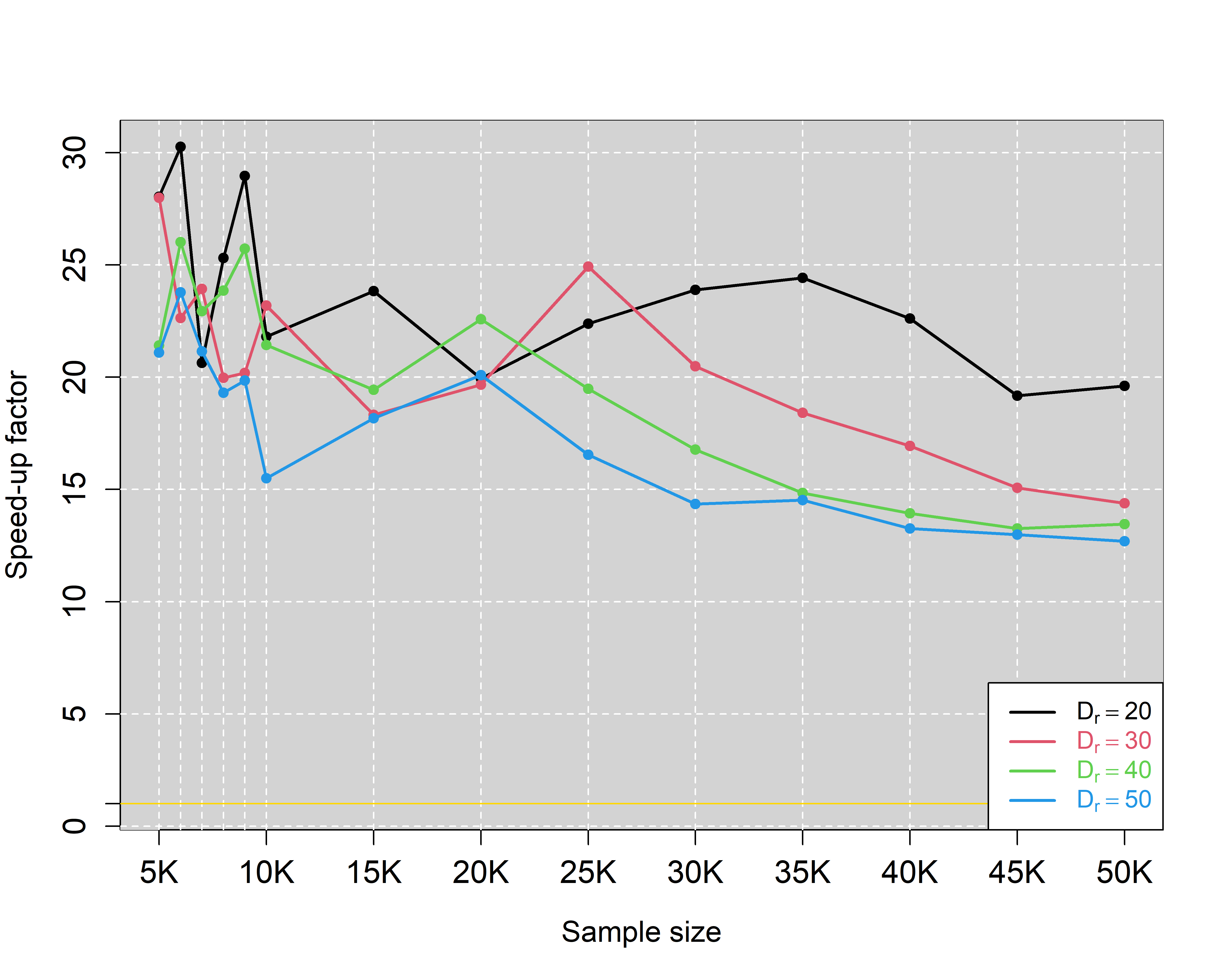}  &
\includegraphics[scale = 0.3]{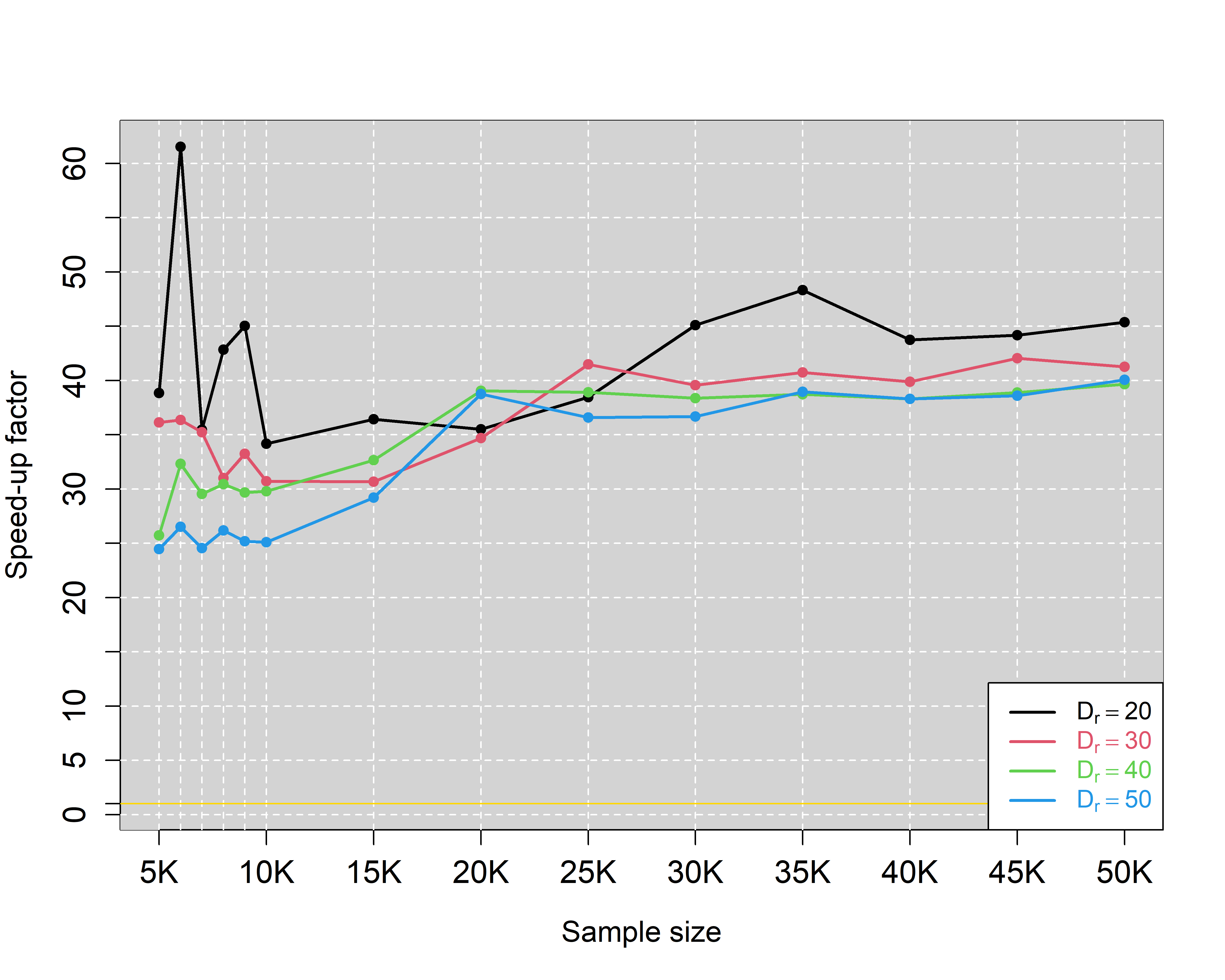}  \\
(e) EM-1 to CIRLS when $p = 15$  &  (f) EM-2 to CIRLS when $p=15$  \\ 
\includegraphics[scale = 0.3]{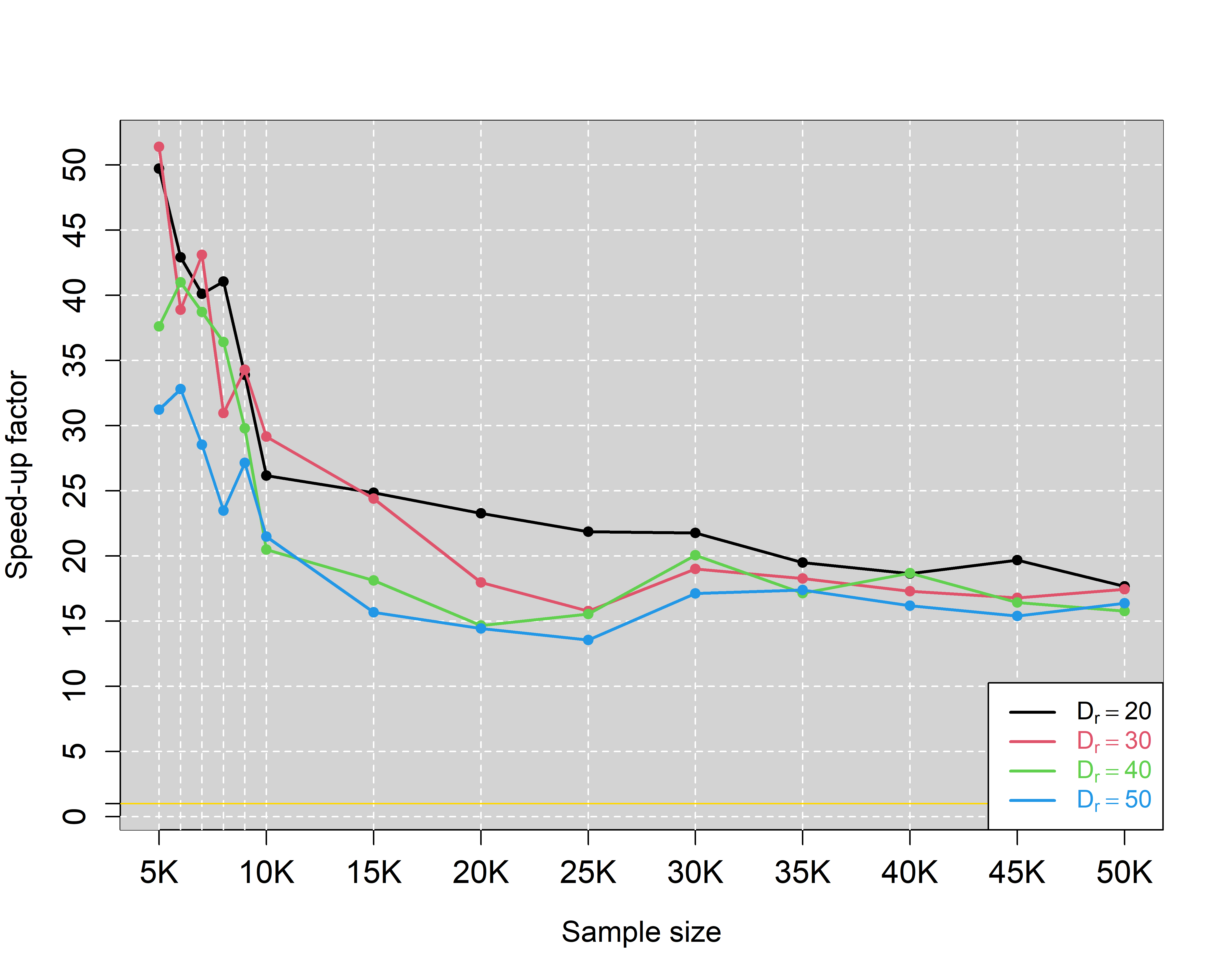}  &
\includegraphics[scale = 0.3]{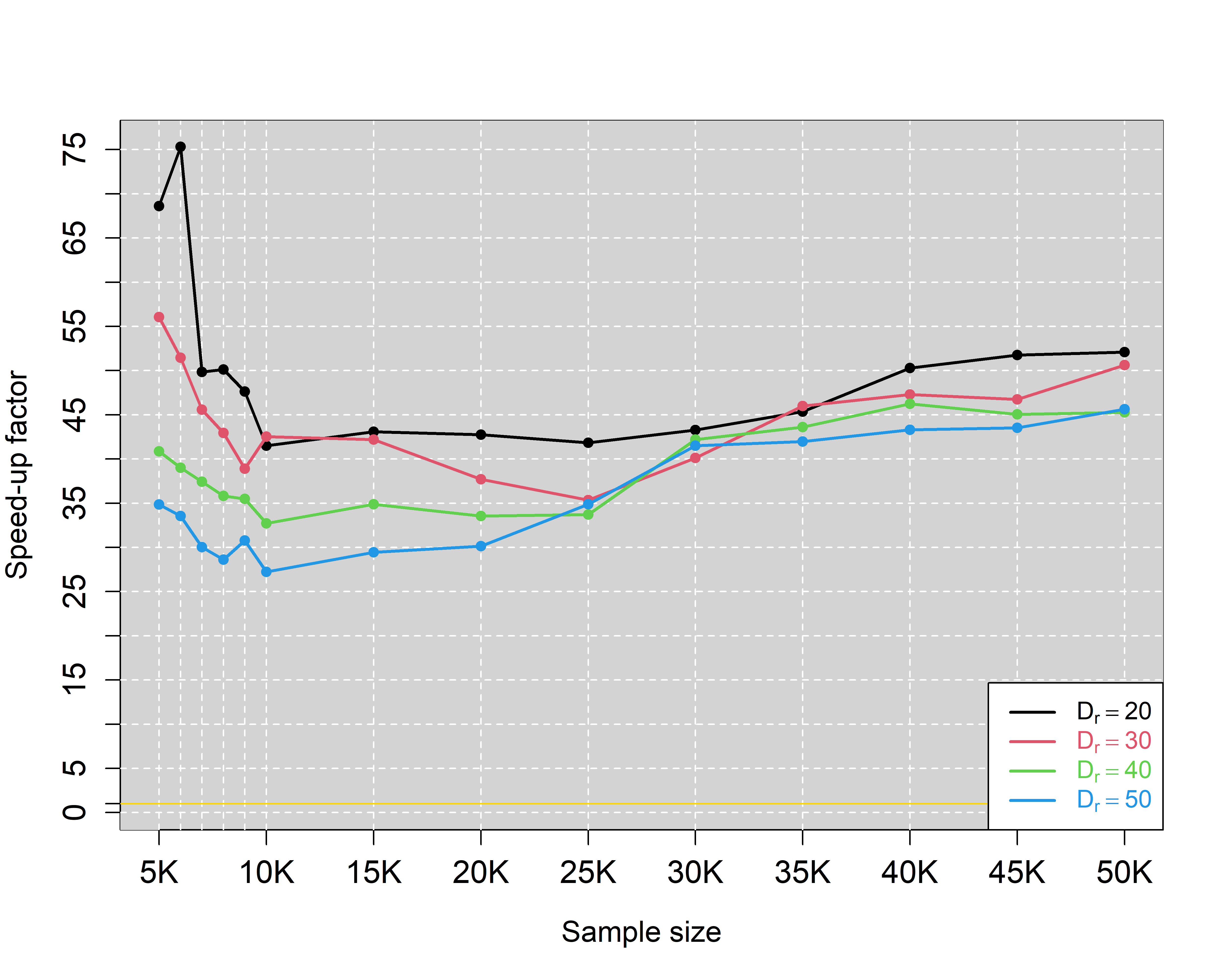}  \\
(g) EM-1 to CIRLS when $p = 20$  &  (h) EM-2 to CIRLS when $p=20$  \\ 
\end{tabular}
\caption{Speed-up factors of the two algorithms, when the two compositions are independent: running time of EM divided by the running time of CIRLS. The horizontal golden line indicates the unity speed-up factor, when the running times are equal. \label{speed1} }
\end{figure}

Figure \ref{speed1} visualizes the speed-up factors of CIRLS to EM when the two compositions are independent. When $p=5$, the speed-up factors of CIRLS compared to EM-1 range from 6 up to 99, whereas compared to EM-2 they range from 6 up to 65. When $p=10$, the EM-1 is 20--157 times slower, whereas EM-2 is 13--76 times slower. When $p=15$ EM-1 is 13--30 times slower than CIRLS, and EM-2 is 24--62 times slower. Finally, when $p=20$ EM-1 is 14--51 times slower than CIRLS, and EM-2 is 27--75 times slower. Regarding the KLD, the discrepancy between EM and CIRLS was $\approx 10^{-7}-10^{-6}$. 

\begin{figure}[!ht]
\centering
\begin{tabular}{cc}
\includegraphics[scale = 0.3]{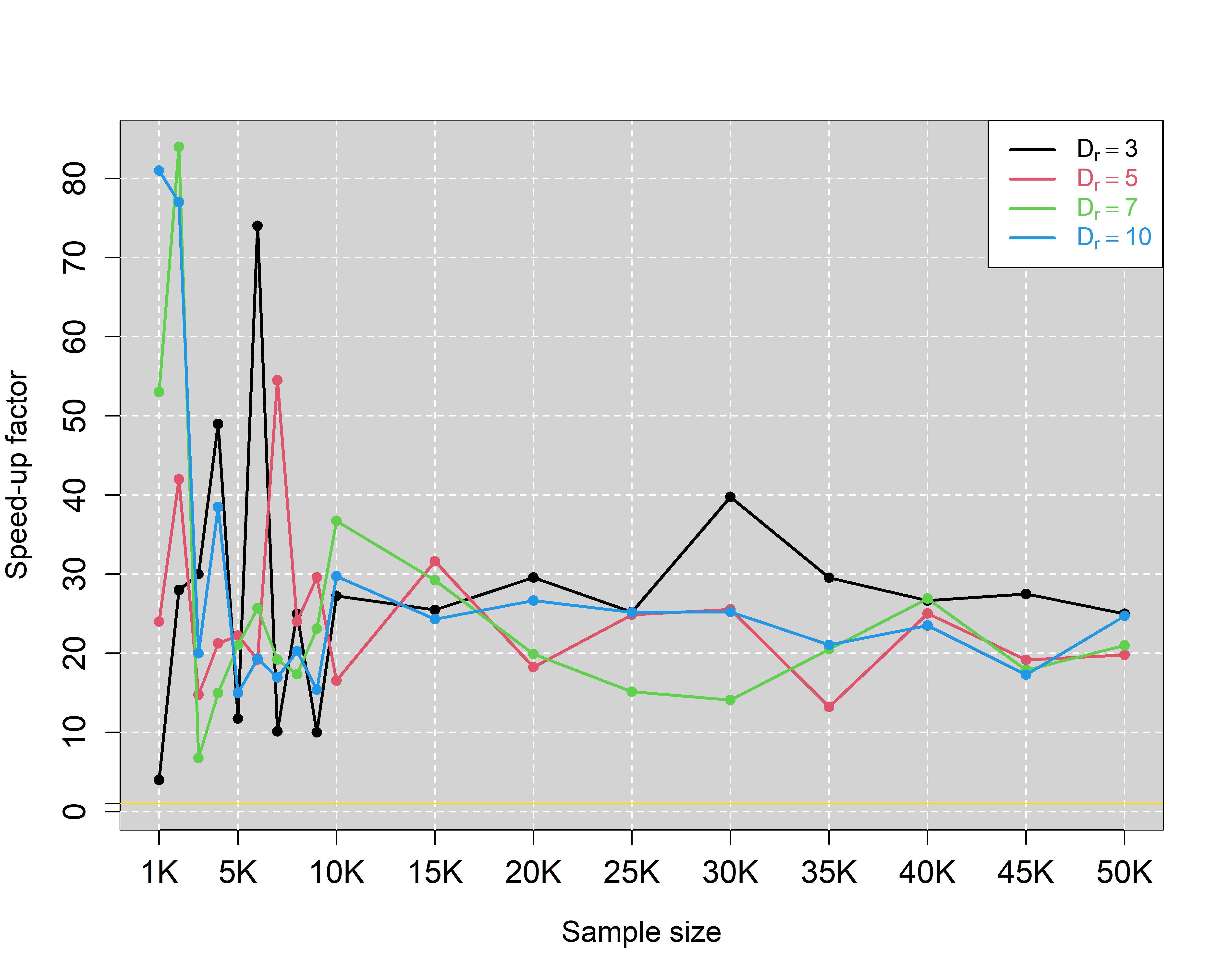}  &
\includegraphics[scale = 0.3]{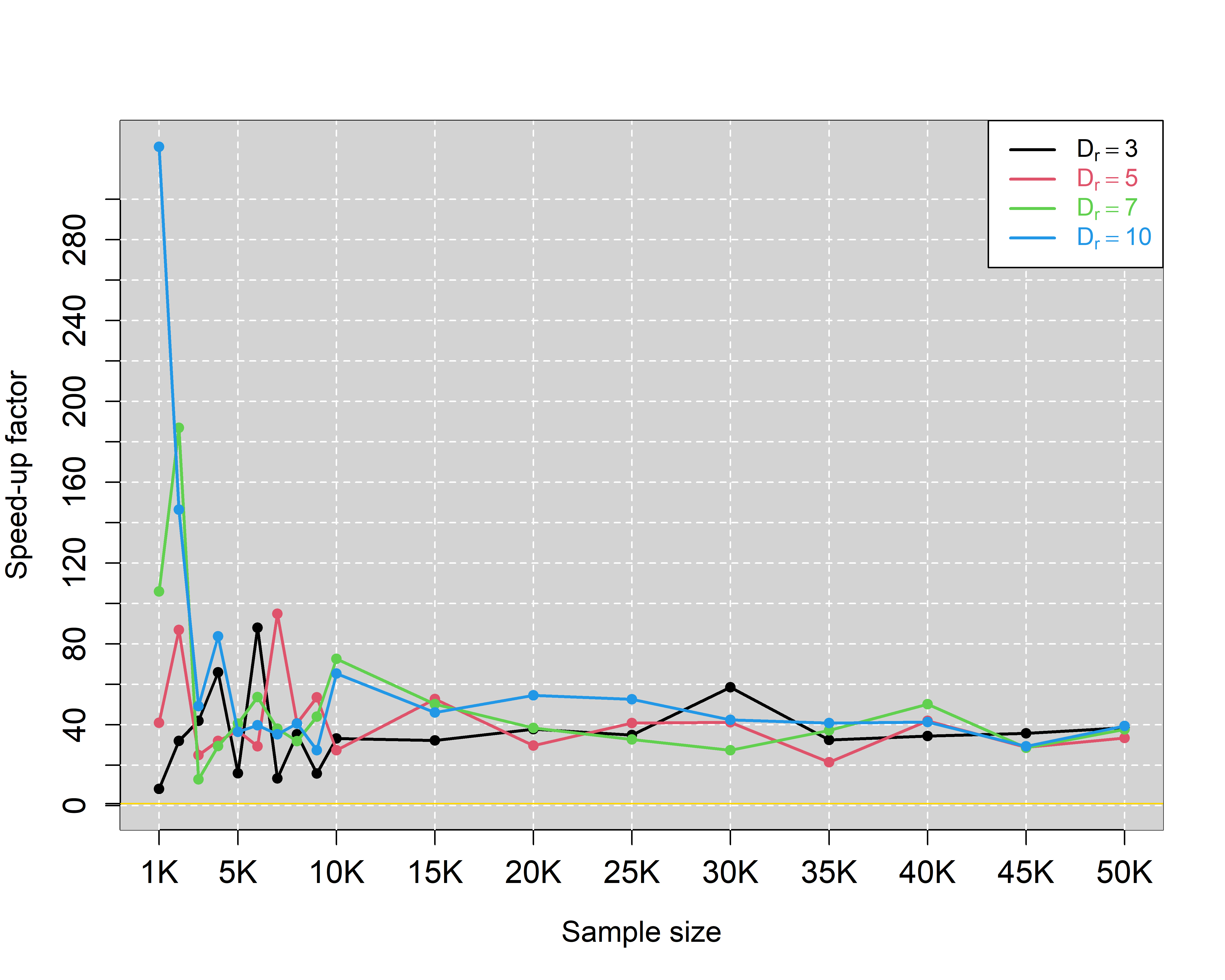}  \\
(a) EM-1 to CIRLS when $p = 5$  &  (b) EM-2 to CIRLS when $p=5$ \\
\includegraphics[scale = 0.3]{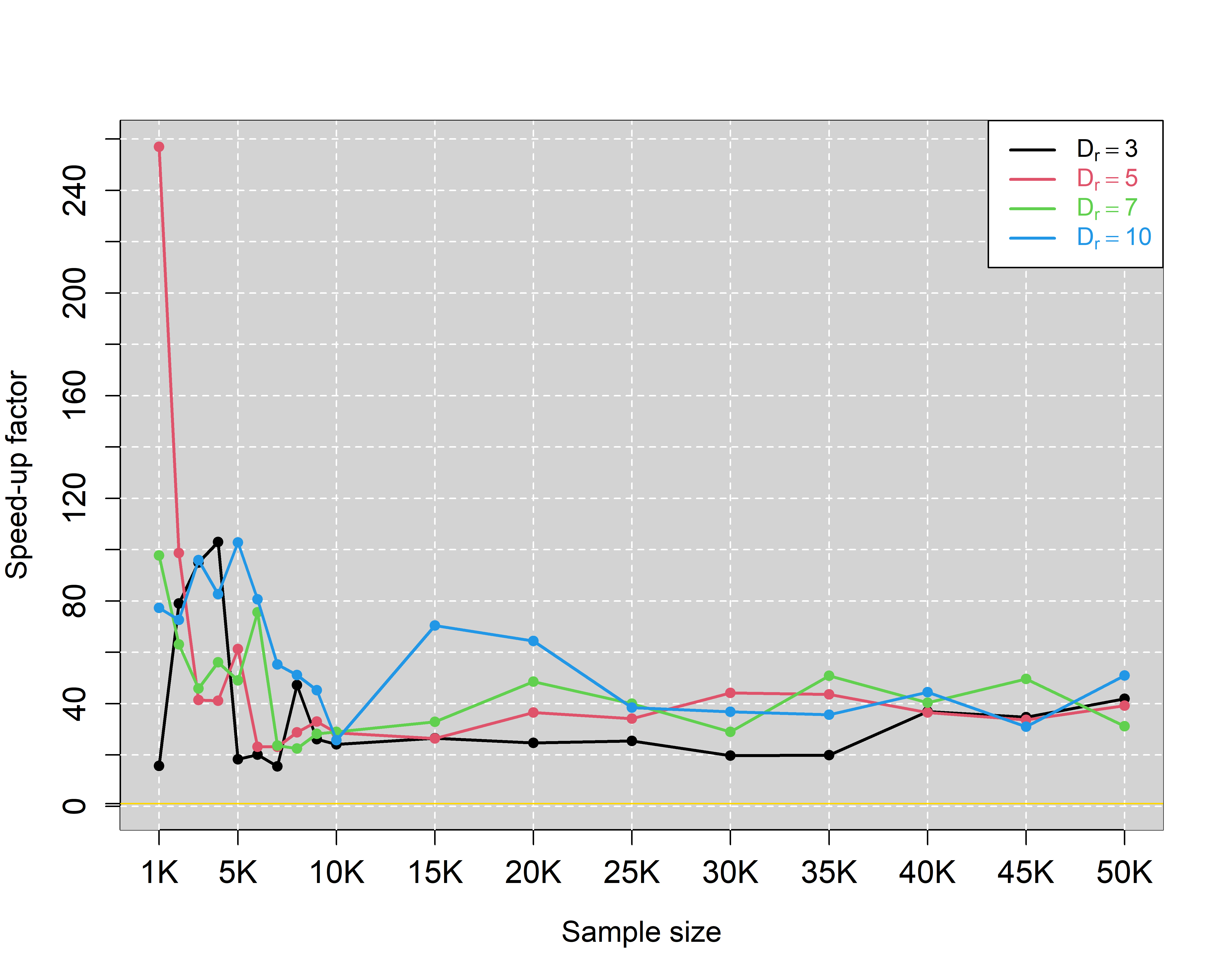}  &
\includegraphics[scale = 0.3]{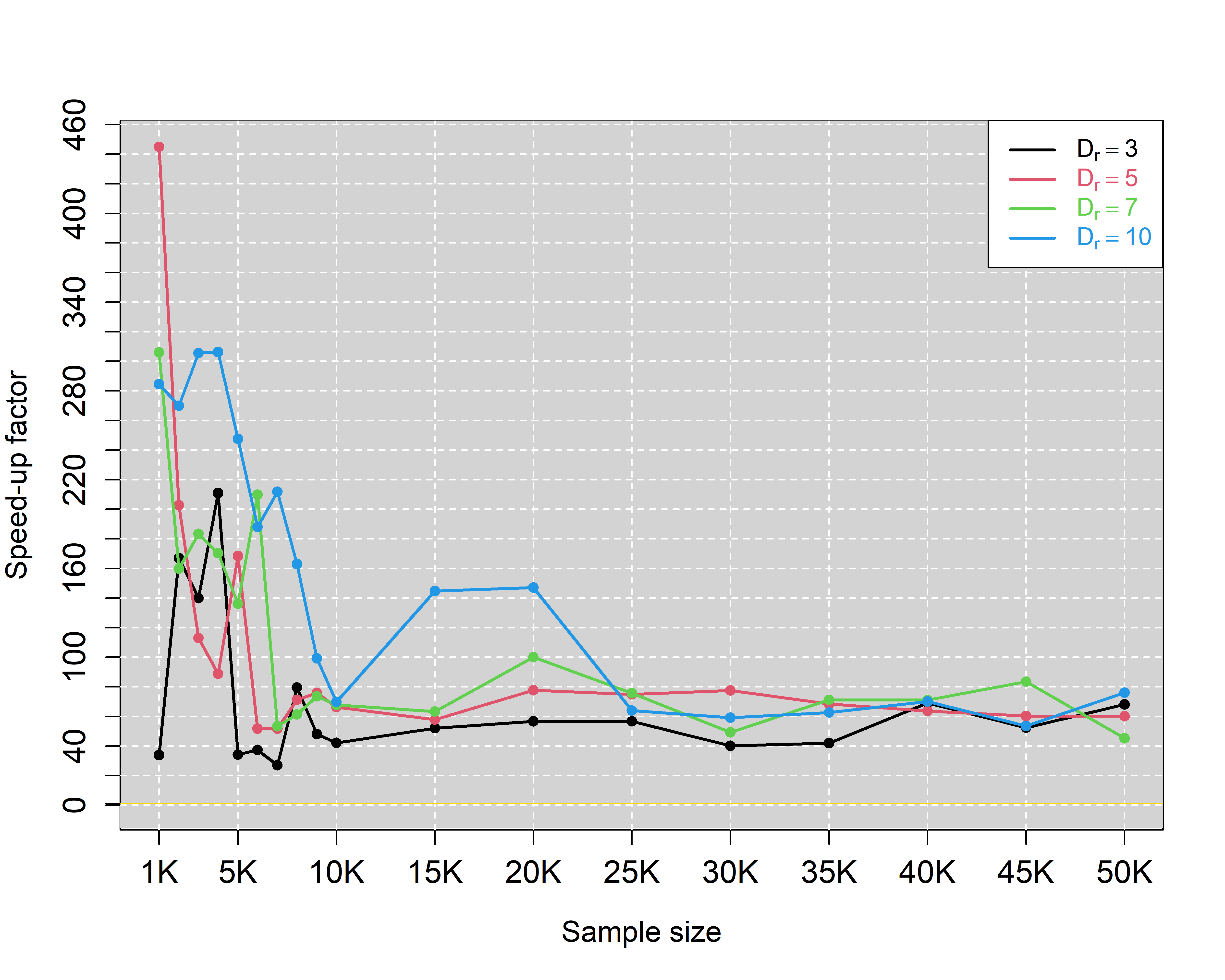}  \\
(c) EM-1 to CIRLS when $p = 10$  &  (d) EM-2 to CIRLS when $p=10$ \\
\includegraphics[scale = 0.3]{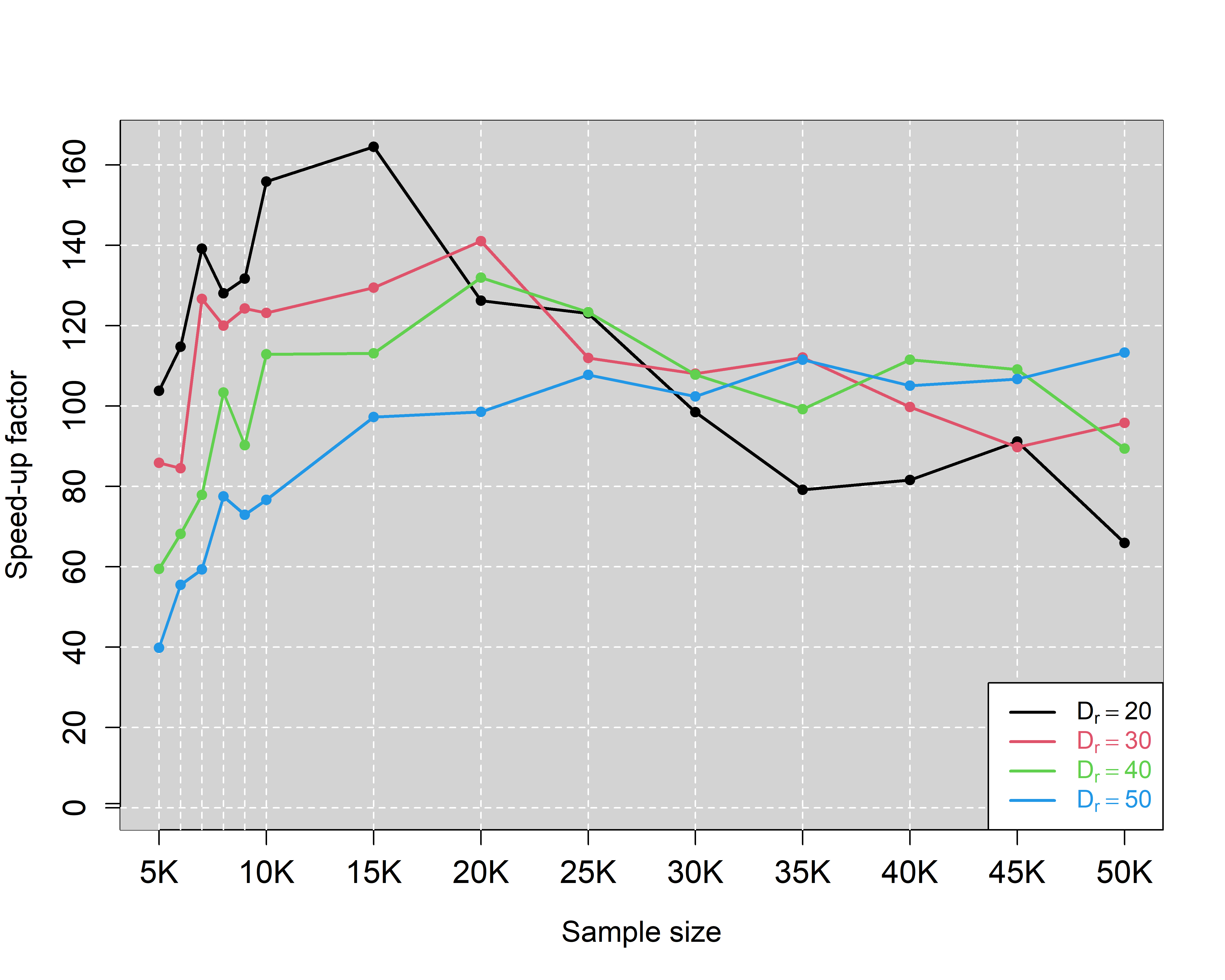}  &
\includegraphics[scale = 0.3]{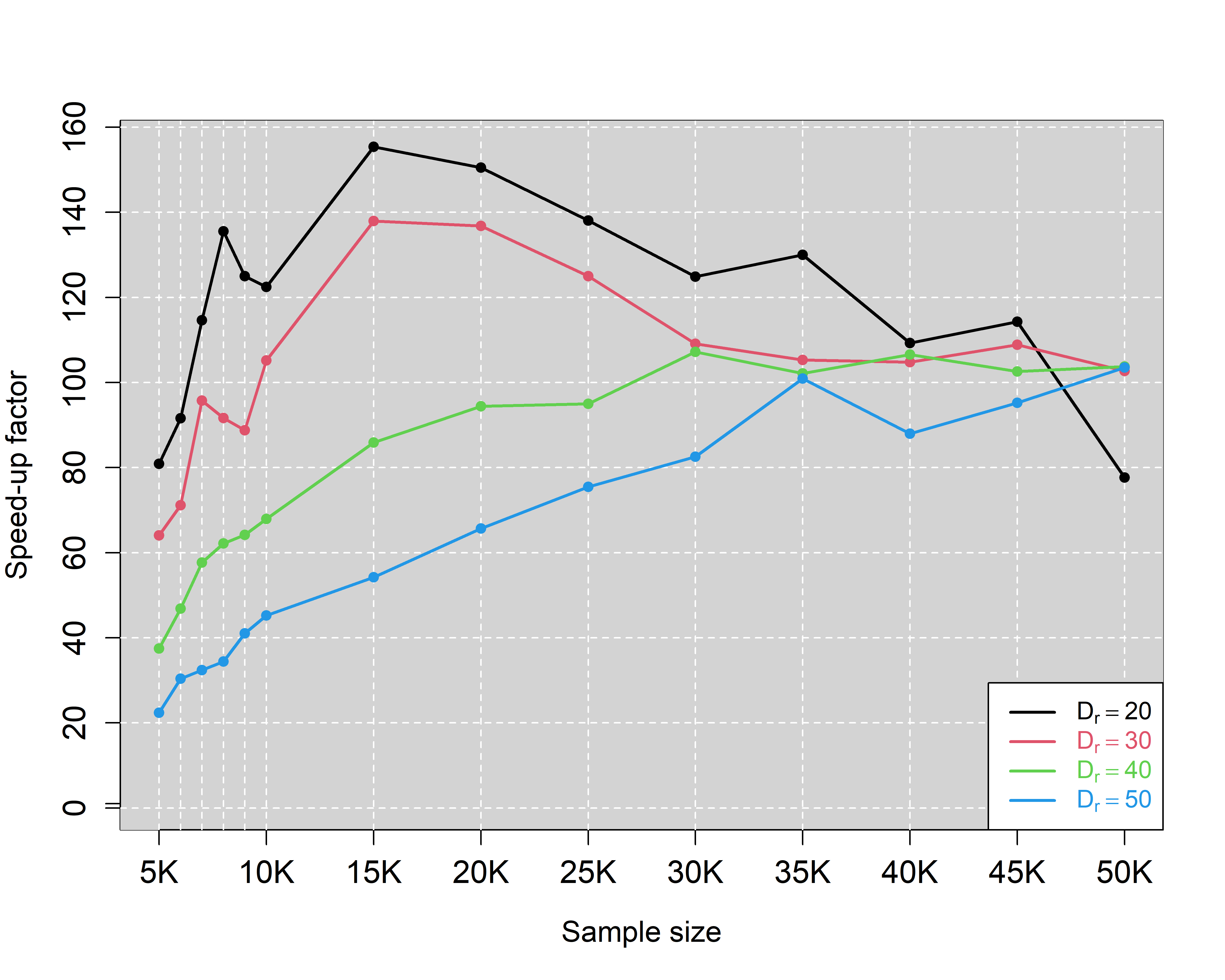}  \\
(e) EM-1 to CIRLS when $p = 15$  &  (f) EM-1 to CIRLS when $p=20$ \\
\end{tabular}
\caption{Speed-up factors of the two algorithms, when the two compositions are dependent: running time of EM divided by the running time of CIRLS. The horizontal golden line indicates the unity speed-up factor, when the running times are equal. \label{speed2} }
\end{figure}

Figure \ref{speed2} visualizes the speed-up factors when the compositions are dependent. When $p=5$, the speed-up factors of CIRLS compared to EM-1 range from 4 up to 84, whereas compared to EM-2 they range from 8 up to 326. When $p=10$, the EM-1 is 15 up 257 times slower, whereas EM-2 is 27-445 times slower. With $p=15$ and $p=20$, EM-1 becomes extremely slow, slower than EM-2, due to large memory requirements and the speed-up factors are not presented. EM-2 is 40-165 times slower than CIRLS when $p=15$, and 22-155 times slower than CIRLS when $p=20$. Notably, there was a discrepancy in the produced KLD values. CIRLS always produced slightly higher KLD than the EM algorithm, the discrepancy was $\approx 10^{-5}-10^{-4}$. 

\subsection{Scalability of the EM and CIRLS algorithms}
To further examine and compare the scalability of the two algorithms, we estimated their scalability rate by fitting the following model $t=\alpha n^\beta$, or $\log(t)=\alpha+\beta \log(n)$ \citep{goldsmith2007}, and report the $\beta$ coefficient, where $t$ and $n$ denote the running time and sample size. Tables \ref{rat} and \ref{rat2} present the empirically estimated scalability rates of both algorithms. Notably, all implementations exhibit linear or sublinear scalability rates, depending on the dimensionalities of the compositions.

\begin{table}[ht]
\centering
\caption{Estimated scalability rates of the EM and CIRLS algorithms when the responses and the predictor variables are independent.}
\label{rat}
\begin{tabular}{r|l|rrrr}
\toprule
& Algorithm    & $D_r=3$  & $D_r=5$  & $D_r=7$  & $D_r=10$ \\  \midrule
$p=5$  & EM-1  & 0.89 & 0.90 & 0.89 & 0.96 \\ 
       & EM-2  & 0.77 & 0.78 & 0.77 & 0.83 \\ 
       & CIRLS & 0.74 & 0.92 & 0.93 & 0.86 \\  \midrule
$p=10$ & EM-1  & 0.76 & 0.78 & 0.82 & 0.76 \\  
       & EM-2  & 0.79 & 0.75 & 0.75 & 0.71 \\ 
       & CIRLS & 1.01 & 1.02 & 0.82 & 0.92 \\ \midrule
$p=15$ & EM-1  & 0.64 & 0.54 & 0.41 & 0.34 \\ 
       & EM-2  & 0.64 & 0.54 & 0.41 & 0.34 \\ 
       & CIRLS & 0.75 & 0.71 & 0.67 & 0.57 \\  \midrule
$p=20$ & EM-1  & 0.30 & 0.18 & 0.14 & 0.13 \\ 
       & EM-2  & 0.62 & 0.62 & 0.63 & 0.62 \\ 
       & CIRLS & 0.72 & 0.65 & 0.55 & 0.45 \\ 
\bottomrule
\end{tabular}
\end{table}

\begin{table}[ht]
\centering
\caption{Estimated scalability rates of the EM and CIRLS algorithms when the responses and the predictor variables are dependent.}
\label{rat2}
\begin{tabular}{r|l|rrrr}
\toprule
& Algorithm    & $D_r=3$  & $D_r=5$  & $D_r=7$  & $D_r=10$ \\  \midrule
$p=5$  & EM-1  & 0.92 & 0.91 & 0.83 & 0.66 \\ 
       & EM-2  & 0.96 & 0.93 & 0.86 & 0.79 \\ 
       & CIRLS & 0.74 & 1.01 & 1.01 & 1.01 \\ \midrule
$p=10$ & EM-1  & 0.80 & 0.63 & 0.44 & 0.40 \\ 
       & EM-2  & 0.81 & 0.71 & 0.66 & 0.66 \\ 
       & CIRLS & 0.91 & 0.99 & 0.82 & 0.89 \\ \midrule
$p=15$ & EM-1  & 0.69 & 0.68 & 0.65 & 0.64 \\ 
       & CIRLS & 0.91 & 0.70 & 0.48 & 0.29 \\ \midrule
$p=20$ & EM-1  & 0.91 & 0.79 & 0.80 & 0.80 \\ 
       & CIRLS & 0.88 & 0.63 & 0.40 & 0.18 \\ 
\bottomrule
\end{tabular}
\end{table}

\subsection{Theoretical justification for KLD discrepancy}
The small discrepancy between EM and CIRLS arises from CIRLS's use of local quadratic approximations via Taylor expansion. While EM maximizes the exact KLD objective, CIRLS solves a sequence of quadratic programs that approximate the KLD surface around the current iterate. The negligible magnitude of this discrepancy ($\approx 10^{-5} - 10^{-4}$) demonstrates that the second-order Taylor approximation is highly accurate in the TFLR setting, making CIRLS a valid and efficient alternative to EM. This is a well-understood trade-off in optimization: Newton-type methods (like IRLS/CIRLS) converge faster but use approximations, while EM guarantees exact monotonic improvement but requires more iterations.

A critical observation is that the Hessian matrix $\mathbf{H}$ is \textbf{singular} due to the simplex constraints. Specifically, for each row of $\bm{B}$, we have the constraint $\sum_{k=1}^{D_r} B_{jk} = 1$, which means the parameters lie in a $(D_r - 1)$-dimensional subspace. This introduces one zero eigenvalue per row, making the full Hessian rank-deficient. The rank of the Hessian is $\text{rank}(\bm{H}) = D_p \cdot (D_r - 1) < D_p \cdot D_r$. This means $\bm{H}$ has exactly $D_p$ zero eigenvalues (one per simplex constraint).

Both EM and CIRLS solve a constrained optimization problem, for which the Karush-Kuhn-Tucker (KKT) conditions \citep{karush1939minima,gale1951linear} at optimality are
\begin{equation*}
\nabla\ell(\bm{B}^*) + \sum_{j=1}^{D_p} \lambda_j \nabla g_j(\bm{B}^*) + \bm{\nu} = \bm{0},
\end{equation*}
where:
\begin{itemize}
\item $g_j(\bm{B}) = \sum_{k=1}^{D_r} B_{jk} - 1 = 0$ is the $j$-th equality constraint (the simplex constraint for the $j$-th row).
\item $\lambda_j \in \bm{R}$ is the Lagrange multiplier for the $j$-th equality constraint.
\item $\nabla g_j(\bm{B})$ is the gradient of the $j$-th constraint; when $\bm{B}$ is vectorized, this is a vector with 1's in positions corresponding to $B_{j1}, \ldots, B_{jD_r}$ and 0's elsewhere.
\item $\bm{\nu} \geq \bm{0}$ is the vector of Lagrange multipliers for the inequality constraints $B_{jk} \geq 0$.
\end{itemize}

The proper framework is to consider the projected Hessian $\bm{H}_{\text{proj}}$ onto the tangent space of the constraints. For the simplex constraints, this is the Hessian restricted to the $(D_r - 1)$-dimensional subspace for each component. Let $\bm{P}$ be the projection matrix onto the constraint manifold. Then:
\begin{equation*}
\bm{H}_{\text{proj}} = \bm{P}^T \bm{H} \bm{P}.
\end{equation*}

For problems involving simplicial data, $\bm{H}_{\text{proj}}$ is typically positive definite with minimum eigenvalue $\lambda_{\min}^{\text{proj}} > 0$ in the constraint subspace. For two feasible points (matrices) $\bm{B}_1$ and $\bm{B}_2$ (both satisfying the simplex constraints), the strong convexity in the constraint subspace gives
\begin{equation*}
\ell(\bm{B}_1) - \ell(\bm{B}_2) \geq \nabla\ell(\bm{B}_2)^\top(\bm{B}_1 - \bm{B}_2) + \frac{\lambda_{\min}^{\text{proj}}}{2}\|\bm{P}(\bm{B}_1 - \bm{B}_2)\|^2.
\end{equation*}

Since both solutions are feasible, $\bm{P}(\bm{B}_1 - \bm{B}_2) = \bm{B}_1 - \bm{B}_2$ (the difference vector is already in the constraint subspace). Let $\Delta\ell = \ell(\bm{B}^*_{\text{EM}}) - \ell(\bm{B}^*_{\text{CIRLS}})$ be the observed KLD difference. Then:
\begin{equation*}
\|\bm{B}^*_{\text{EM}} - \bm{B}^*_{\text{CIRLS}}\| \leq \sqrt{\frac{2|\Delta\ell|}{\lambda_{\min}^{\text{proj}}}}.
\end{equation*}

For the TFLR model, the projected Hessian can be approximated by
\begin{equation*}
\bm{H}_{\text{proj}} \approx \text{diag}\left(\bm{X}^\top \bm{V}_1^{\text{proj}} \bm{X}, \ldots, \bm{X}^\top \mathbf{V}_{D_r-1}^{\text{proj}} \bm{X}\right),
\end{equation*}
where $\bm{V}_k^{\text{proj}}$ are the projected weight matrices, and the minimum eigenvalue is aprproximately given by
\begin{equation*}
\lambda_{\min}^{\text{proj}} \approx \mathcal{O}\left(n \cdot \min_{i,j,k} \left[\frac{x_{ij}^2 \cdot y_{ik}(1-y_{ik})}{(\sum_j x_{ij}B_{jk})^2}\right]\right).
\end{equation*}

For well-conditioned simplicial data (no components too close to 0 or 1), this is approximately $\lambda_{\min}^{\text{proj}} \approx \mathcal{O}(n)$. Given the observed KLD differences of $|\Delta\ell| \approx 10^{-5} - 10^{-4}$ in the dependent case, and using $\lambda_{\min}^{\text{proj}} \approx c \cdot n$ where $c$ is a problem-dependent constant (typically $c \in [0.1, 1]$)
\begin{equation*}
\|\bm{B}^*_{\text{EM}} - \bm{B}^*_{\text{CIRLS}}\|_F \lesssim \mathcal{O}\left(\sqrt{\frac{10^{-4}}{c \cdot n}}\right)
\end{equation*}

For example, for $n = 1,000$ and assuming $c \approx 0.5$ we obtain
\begin{equation*}
\|\bm{B}^*_{\text{EM}} - \bm{B}^*_{\text{CIRLS}}\|_F \lesssim \mathcal{O}\left(\sqrt{\frac{10^{-4}}{500}}\right) \approx 4.5 \times 10^{-4}
\end{equation*}

\subsection{Permutation testing}
We examined the effect of this discrepancy on the permutation testing. Using the same simulation studies described earlier, we performed permutation testinf for independence between the simplicial responses and the simplicial predictors. We performed 100 iterations and for each iteration we computed the p-value based on the EM algorithm and based on the CIRLS algorithm using 300 permutations. For each iteration, we computed the correlation coefficient between the permuted KLD values of the two algorithms. The correlation coefficient was always equal to 1 and the p-values were always the same. 

\clearpage
\section{Conclusions}
We adapted the IRLS methodology employed in generalized linear models to the TFLR framework for simplicial-simplicial regression. This adaptation exploits the formal equivalence between TFLR and logistic regression with simplicial constraints imposed on the coefficient matrix. Irrespective of the dimensionalities of both response and predictor simplices, this approach yields substantial computational improvements, with speed-up factors ranging from $6\times$ to over $300\times$. Both algorithms exhibit linear or sublinear scalability with respect to sample size and dimensionality. However, the exceptional computational efficiency of CIRLS entails a modest trade-off: when compositions exhibit dependence structure, CIRLS attains KLD values that deviate marginally ($\approx 10^{-6}$--$10^{-4}$) from the global minimum. 

This discrepancy is negligible for most practical applications, particularly for permutation-based and bootstrap hypothesis testing procedures. Our simulation studies demonstrate that permutation-based $p$-values derived from EM and CIRLS exhibit strong concordance. 

The reduced computational burden of CIRLS confers substantial advantages in these contexts, especially when analyzing large-scale data (with respect to sample size, dimensionality, or both). These benefits extend to scenarios involving multiple datasets or extensive simulation studies, where computational efficiency becomes paramount. Moreover, CIRLS is naturally amenable to parallelization, and through judicious programming strategies (e.g., pre-computation of invariant quantities), permutation and bootstrap testing procedures can achieve exceptional computational efficiency. 

The CIRLS framework enables several promising methodological extensions in simplicial-simplicial regression. \cite{fiksel2022} proposed minimizing the KLD from observed to fitted compositions, while \cite{tsagris2025} explored objective functions based on $L_2$ and $L_1$ norms. $\phi$-divergence regression represents another natural extension, encompassing the symmetric KLD and Jensen-Shannon divergence (a symmetrized variant of the KLD), among others. The CIRLS algorithm can provide approximate solutions to such regression problems, yielding computationally efficient initial values for subsequent refinement via the EM algorithm.

\subsection*{Declarations}
\textbf{Conflict of interest statement}: The author declares no conflict of interest. \\
\textbf{Funding Information}: This research received no funding. \\
\textbf{Data availability}: This research has used no data.

\bibliographystyle{apalike}
\bibliography{vivlio}

\end{document}